\def\astropbibitem{\@lbibitem}

\def\@lbibitem#1#2#3{\item[\hfill]\if@filesw
 { \def\protect##1{\string ##1\space}\immediate
 \write\@auxout{\string\astropbibcite{#3}{#1}{#2}}\fi\ignorespaces}}

\def\astropbibcite#1#2#3{\global\@namedef{b@#1}{#2\ #3}
 \global\@namedef{newb@#1}{#2\ (#3} \global\@namedef{nameb@#1}{#2}
\global\@namedef{yearb@#1}{#3}}

\let\citation\@gobble
\def\cite{\@ifnextchar [{\@tempswatrue\@citex}{\@tempswafalse\@citex[]}}
\def\citeone{\@ifnextchar [{\@tempswatrue\@newcitex}
			   {\@tempswafalse\@newcitex[]}}
\def\citename{\@ifnextchar [{\@tempswatrue\@namecitex}
			   {\@tempswafalse\@namecitex[]}}
\def\citeyear{\@ifnextchar [{\@tempswatrue\@yearcitex}
			   {\@tempswafalse\@yearcitex[]}}
\def\@citex[#1]#2{\if@filesw\immediate\write\@auxout{\string\citation{#2}}\fi
  \def\@citea{}\@cite{\@for\@citeb:=#2\do
     {\@citea\def\@citea{; }\@ifundefined
       {b@\@citeb}{{\bf ?}\@warning
       {Citation `\@citeb' on page \thepage \space undefined}}
{\csname b@\@citeb\endcsname}}}{#1}}
\def\@newcitex[#1]#2{\if@filesw\immediate\write\@auxout{\string\citation{#2}}
\fi
  \def\@newcitea{}\@newcite{\@for\@newciteb:=#2\do
     {\@newcitea\def\@newcitea{; }\@ifundefined
       {newb@\@newciteb}{{\bf ? (?}\@warning
       {Citation `\@newciteb' on page \thepage \space undefined}}
{\csname newb@\@newciteb\endcsname}}}{#1}}

\def\@namecitex[#1]#2{\if@filesw\immediate\write\@auxout{\string\citation{#2}}
\fi
  \def\@namecitea{}\@namecite{\@for\@nameciteb:=#2\do
     {\@namecitea\def\@namecitea{; }\@ifundefined
       {nameb@\@nameciteb}{{\bf ? (?}\@warning
       {Citation `\@nameciteb' on page \thepage \space undefined}}
{\csname nameb@\@nameciteb\endcsname}}}{#1}}

\def\@yearcitex[#1]#2{\if@filesw\immediate\write\@auxout{\string\citation{#2}}
\fi
  \def\@yearcitea{}\@yearcite{\@for\@yearciteb:=#2\do
     {\@yearcitea\def\@yearcitea{; }\@ifundefined
       {yearb@\@yearciteb}{{\bf ? (?}\@warning
       {Citation `\@yearciteb' on page \thepage \space undefined}}
{\csname yearb@\@yearciteb\endcsname}}}{#1}}
\let\bibdata=\@gobble
\let\bibstyle=\@gobble
\def\bibliography#1{\if@filesw\immediate\write\@auxout{\string\bibdata{#1}}\fi
  \@input{\jobname.bbl}}
\def\bibliographystyle#1{\if@filesw\immediate\write\@auxout
    {\string\bibstyle{#1}}\fi}
\def\nocite#1{\@bsphack
  \if@filesw\immediate\write\@auxout{\string\citation{#1}}\fi
  \@esphack}
\def\@cite#1#2{({#1\if@tempswa , #2\fi})}
\def\@newcite#1#2{{#1\if@tempswa , #2\fi})}
\def\@namecite#1#2{#1}
\def\@yearcite#1#2{#1}
\newcommand{\citebare}[1]{{\citename{#1}\ \citeyear{#1}}}

\def\thebibliography#1{\section*{References}\list
 {}{\setlength\labelwidth{1.4em}\leftmargin\labelwidth
 \setlength\parsep{0pt}\setlength\itemsep{0pt}
 \setlength{\itemindent}{-\leftmargin}
 \usecounter{enumi}}
 \def\newblock{\hskip .11em plus .33em minus -.07em}
 \sloppy
 \sfcode`\.=1000\relax}

\def~{\penalty\@M\kern3pt}

\def\@ptsize{0} \@namedef{ds@11pt}{\def\@ptsize{1}}
\@namedef{ds@12pt}{\def\@ptsize{2}}
\def\ds@twoside{\@twosidetrue \@mparswitchtrue}
\def\ds@draft{\overfullrule 5\p@}
\def\ds@noframe{\let\picplace\vspace}
\newif\if@referee \@refereefalse
\def\ds@referee{\@refereetrue}

\def\picplace#1{\vbox{\hrule\@height 0.4pt\@width\hsize
\hbox to\hsize{\vrule\@width 0.4pt\@height#1\hfil
\vrule\@width 0.4pt\@height#1}\hrule\@height 0.4pt\@width\hsize}}

\@options
\ds@twoside

\def\thisbottomragged{\def\@textbottom{\vskip\z@ plus.0001fil
\global\let\@textbottom\relax}}

\def\@normalsize{\@setsize\normalsize{12pt}\xpt\@xpt
\abovedisplayskip=3 mm plus6pt minus 4pt
\belowdisplayskip \abovedisplayskip
\abovedisplayshortskip=0mm plus6pt
\belowdisplayshortskip=2 mm plus4pt minus 4pt
\def\@listi{\topsep 4pt plus 2pt minus 2pt
\leftmargin \leftmargini
\parsep 0pt
\itemsep \parsep}}
\def\small{\@setsize\small{11pt}\ixpt\@ixpt
\abovedisplayskip 8.5pt plus 3pt minus 4pt
\belowdisplayskip \abovedisplayskip
\abovedisplayshortskip \z@ plus 2pt
\belowdisplayshortskip 4pt plus2pt minus 2pt
\def\@listi{\topsep 3pt plus 2pt minus 2pt
\leftmargin \leftmargini
\parsep 0pt
\itemsep \parsep}}

\def\scriptsize{\@setsize\scriptsize{8pt}\viipt\@viipt}
\def\tiny{\@setsize\tiny{6pt}\vpt\@vpt}
\def\large{\@setsize\large{17pt}\xivpt\@xivpt}
\def\Large{\@setsize\Large{17pt}\xivpt\@xivpt}
\def\LARGE{\@setsize\LARGE{20pt}\xviipt\@xviipt}
\def\huge{\@setsize\huge{25pt}\xviipt\@xviipt}
\def\Huge{\@setsize\Huge{30pt}\xviipt\@xviipt}
\@normalsize\rm

\leftmargin    \parindent
\leftmargini   \leftmargin
\leftmarginii  \leftmargin
\leftmarginiii \leftmargin
\leftmarginiv  \leftmargin
\leftmarginv   \leftmargin
\leftmarginvi  \leftmargin
\labelwidth\leftmargini
\advance\labelwidth-\labelsep
\def\@listii{\leftmargin\leftmarginii
 \labelwidth\leftmarginii\advance\labelwidth-\labelsep
 \topsep \z@}
\def\@listiii{\leftmargin\leftmarginiii
 \labelwidth\leftmarginiii\advance\labelwidth-\labelsep
 \topsep \z@}
\def\@listiv{\leftmargin\leftmarginiv
 \labelwidth\leftmarginiv\advance\labelwidth-\labelsep}
\def\@listv{\leftmargin\leftmarginv
 \labelwidth\leftmarginv\advance\labelwidth-\labelsep}
\def\@listvi{\leftmargin\leftmarginvi
 \labelwidth\leftmarginvi\advance\labelwidth-\labelsep}

\skip\footins 9pt plus 4pt minus 2pt
-\@lowpenalty \@endparpenalty -\@lowpenalty \@itempenalty -\@lowpenalty

\def\vec#1{\ifmmode\mathchoice{\mbox{\boldmath$\displaystyle#1$}}
{\mbox{\boldmath$\textstyle#1$}}
{\mbox{\boldmath$\scriptstyle#1$}}
{\mbox{\boldmath$\scriptscriptstyle#1$}}\else
\hbox{\boldmath$\textstyle#1$}\fi}

\def\tens#1{\ifmmode\mathchoice{\mbox{$\sf\displaystyle#1$}}
{\mbox{$\sf\textstyle#1$}}
{\mbox{$\sf\scriptstyle#1$}}
{\mbox{$\sf\scriptscriptstyle#1$}}\else
\hbox{$\sf\textstyle#1$}\fi}

\def\part{\par \addvspace{4ex} \@afterindentfalse \secdef\@part\@spart}
\def\@part[#1]#2{\ifnum \c@secnumdepth >\m@ne \refstepcounter{part}
\addcontentsline{toc}{part}{\thepart \hspace{1em}#1}\else
\addcontentsline{toc}{part}{#1}\fi { \parindent 0pt \raggedright \ifnum
\c@secnumdepth >\m@ne \Large \bf Part \thepart \par \nobreak \fi \huge
\bf #2\par } \nobreak \vskip 3ex \@afterheading }
\def\@spart#1{{\parindent 0pt \raggedright \huge \bf #1\par} \nobreak
\vskip 3ex \@afterheading }

\def\section{\@startsection {section}{1}{\z@}{-3.5ex plus -1ex minus
 -.2ex}{1.5ex plus .2ex}{\normalsize\bf\boldmath}}
\def\subsection{\@startsection{subsection}{2}{\z@}{-3.25ex plus -1ex
minus -.2ex}{1.5ex plus .2ex}{\normalsize\it}}
\def\subsubsection{\@startsection{subsubsection}{3}{\z@}{-3.25ex plus
-1ex minus -.2ex}{1.5ex plus .2ex}{\normalsize}}
\def\paragraph{\@startsection {paragraph}{4}{\z@}{3.25ex plus 1ex minus
.2ex}{-.6em}{\normalsize\it}}
\def\subparagraph#1{\typeout{AandA Warning: You should not use
\protect\subparagraph \space in this style.}\vskip0.5cm
You should not use $\backslash${\tt subparagraph} in this
style.\vskip0.5cm}

\def\@sect#1#2#3#4#5#6[#7]#8{\ifnum #2>\c@secnumdepth
     \def\@svsec{}\else
     \refstepcounter{#1}\edef\@svsec{\csname the#1\endcsname.\ }\fi
     \@tempskipa #5\relax
      \ifdim \@tempskipa>\z@
 \begingroup #6\relax
   \@hangfrom{\hskip #3\relax\@svsec}{\interlinepenalty \@M #8\par}
 \endgroup
       \csname #1mark\endcsname{#7}\addcontentsline
  {toc}{#1}{\ifnum #2>\c@secnumdepth \else
        \protect\numberline{\csname the#1\endcsname.}\fi
      #7}\else
 \def\@svsechd{#6\hskip #3\@svsec #8\csname #1mark\endcsname
        {#7}\addcontentsline
      {toc}{#1}{\ifnum #2>\c@secnumdepth \else
        \protect\numberline{\csname the#1\endcsname.}\fi
         #7}}\fi
     \@xsect{#5}}

\def\@xsect#1{\@tempskipa #1\relax
      \ifdim \@tempskipa>\z@
       \par \nobreak
       \addvspace{\@tempskipa}
       \@afterheading
    \else \global\@nobreakfalse \global\@noskipsectrue
       \everypar{\if@noskipsec \global\@noskipsecfalse
     \clubpenalty\@M \hskip -\parindent
     \begingroup \@svsechd \endgroup \unskip
     \hskip -#1
    \else       \clubpenalty \@clubpenalty
      \everypar{}\fi}\fi\ignorespaces}

\def\appendix{\par
 \setcounter{section}{0}
 \setcounter{subsection}{0}
 \def\thesection{\Alph{section}}
 \renewcommand{\theequation}{\thesection\arabic{equation}}
 \setcounter{equation}{0}
 \@addtoreset{equation}{section}}

\def\theenumi{\arabic{enumi}}

\def\theenumii{\alph{enumii}}
\def\p@enumii{\theenumi}

\def\theenumiii{\roman{enumiii}}
\def\p@enumiii{\theenumi(\theenumii)}

\def\p@enumiv{\p@enumiii\theenumiii}

\def\verse{\let\\=\@centercr
 \list{}{\itemsep\z@ \itemindent -1.5em\listparindent \itemindent
 \rightmargin\leftmargin\advance\leftmargin 1.5em}\item[]}

\def\descriptionlabel#1{\hspace\labelsep \bf #1}
\def\description{\list{}{\labelwidth\z@ \itemindent-\leftmargin
 \let\makelabel\descriptionlabel}}

\def\theequation{\arabic{equation}}

\def\titlepage{\@restonecolfalse\if@twocolumn\@restonecoltrue\onecolumn
 \else \newpage \fi \thispagestyle{empty}\c@page\z@}
\def\endtitlepage{\if@restonecol\twocolumn \else \newpage \fi}

\tabbingsep \labelsep

\skip\@mpfootins = \skip\footins
\newcounter{part}
\newcounter {section}
\newcounter {subsection}[section]
\newcounter {subsubsection}[subsection]
\newcounter {paragraph}[subsubsection]
\newcounter {subparagraph}[paragraph]

\def\thepart{\Roman{part}} \def\thesection {\arabic{section}}

\def\@pnumwidth{1.55em}
\def\@tocrmarg {2.55em}
\def\@dotsep{4.5}
\def\tableofcontents{\section*{Contents}
 \@starttoc{toc}}
\def\l@part#1#2{\addpenalty{\@secpenalty}
 \addvspace{2.25em plus 1pt} \begingroup
 \@tempdima 3em \parindent \z@ \rightskip \@pnumwidth \parfillskip
-\@pnumwidth
 {\large \bf \leavevmode #1\hfil \hbox to\@pnumwidth{\hss #2}}\par
 \nobreak \endgroup}
\def\l@section#1#2{\addpenalty{\@secpenalty} \addvspace{0.5em plus 1pt}
\@tempdima 1.5em \begingroup
 \parindent \z@ \rightskip \@pnumwidth
 \parfillskip -\@pnumwidth
 \bf \leavevmode #1\hfil \hbox to\@pnumwidth{\hss #2}\par
 \endgroup}
\def\l@subsection{\@dottedtocline{2}{1.5em}{2.3em}}
\def\l@subsubsection{\@dottedtocline{3}{3.8em}{3.2em}}
\def\l@paragraph{\@dottedtocline{4}{7.0em}{4.1em}}
\def\l@subparagraph{\@dottedtocline{5}{10em}{5em}}
\def\listoffigures{\section*{List of Figures}\@starttoc{lof}}
\def\l@figure{\@dottedtocline{1}{1.5em}{2.3em}}
\def\listoftables{\section*{List of Tables}\@starttoc{lot}}
\let\l@table\l@figure

\def\thebibliography#1{\small\section*{References}\list
 {}{\setlength\labelwidth{1.4em}\leftmargin\labelwidth
 \setlength\parsep{0pt}\setlength\itemsep{0pt}
 \setlength{\itemindent}{-\leftmargin}
 \usecounter{enumi}}
 \def\newblock{\hskip .11em plus .33em minus -.07em}
 \sloppy
 \sfcode`\.=1000\relax}

\newif\if@restonecol
\def\theindex{\@restonecoltrue\if@twocolumn\@restonecolfalse\fi
\columnseprule \z@
\columnsep 35pt\twocolumn[\section*{Index}]
 \thispagestyle{plain}\parindent\z@
 \parskip\z@ plus .3pt\relax\let\item\@idxitem}
\def\@idxitem{\par\hangindent 40pt}

\def\endtheindex{\if@restonecol\onecolumn\else\clearpage\fi}

\def\footnoterule{\kern-3\p@
 \hrule width 2cm
 \kern 2.6\p@}

\long\def\@makefntext#1{\parindent 1em\noindent
 \hbox to 1.5em{\hss$^{\@thefnmark}$}#1}

\def\aafigurecaption{figure}

\long\def\@makecaption#1#2{%
\makeatletter
\ifx\@captype\aafigurecaption\vskip5mm\else\vskip10pt\fi
\makeatother
\small
 \setbox\@tempboxa\hbox{{\bf #1} #2}
 \ifdim \wd\@tempboxa >\hsize \unhbox\@tempboxa\par \else \hbox
to\hsize{\box\@tempboxa\hfil}
 \fi
\makeatletter
\ifx\@captype\aafigurecaption\else\vskip6pt\fi
\makeatother}

\newcounter{figure}
\def\thefigure{\@arabic\c@figure}
\def\fps@figure{htbp}
\def\ftype@figure{1}
\def\ext@figure{lof}
\def\fnum@figure{Fig.\ \thefigure.}
\def\figure{\small\rm\@float{figure}}
\let\endfigure\end@float
\@namedef{figure*}{\small\rm\@dblfloat{figure}}
\@namedef{endfigure*}{\end@dblfloat}

\newcounter{table}
\def\thetable{\@arabic\c@table}
\def\fps@table{htbp}
\def\ftype@table{2}
\def\ext@table{lot}
\def\fnum@table{Table \thetable.}
\def\table{\small\rm\@float{table}}
\let\endtable\end@float
\@namedef{table*}{\small\rm\@dblfloat{table}}
\@namedef{endtable*}{\end@dblfloat}

\newcounter{@inst}
\newcounter{@auth}
\newdimen\instindent

\def\institute#1{\gdef\@institute{#1}}

\def\institutename{\par
 \begingroup
 \parindent=0pt
 \parskip=0pt
 \setcounter{@inst}{1}%
 \def\and{\par\stepcounter{@inst}%
 \noindent
 \hbox to\instindent{\hss$^{\the@inst}$\enspace}\ignorespaces}%
 \setbox0=\vbox{\def\thanks##1{}\@institute}
 \ifnum\c@@inst>9\relax\setbox0=\hbox{$^{88}$\enspace}%
                 \else\setbox0=\hbox{$^{8}$\enspace}\fi
 \instindent=\wd0\relax
 \ifnum\c@@inst=1\relax
 \else
   \setcounter{@inst}{1}%
   \noindent
   \hbox to\instindent{\hss$^{\the@inst}$\enspace}\ignorespaces
 \fi
 \ignorespaces
 \@institute\par
 \endgroup}

\def\offprints#1{\begingroup
\def\protect{\noexpand\protect\noexpand}\xdef\@thanks{\@thanks
\protect\footnotetext[0]{\unskip\hskip-1.5em{\it Send offprint requests
to\/}: \ignorespaces#1}}\endgroup\ignorespaces}

\def\@thanks{}

\long\def\@makefntext#1{\parindent 1em\noindent
            \hbox to1.5em{$\m@th^{\@thefnmark}$\hss}#1}

\def\@fnsymbol#1{\ifcase#1\or\star\or{\star\star}\or{\star\star\star}%
   \or \dagger\or \ddagger\or
   \mathchar "278\or \mathchar "27B\or \|\or **\or \dagger\dagger
   \or \ddagger\ddagger \else\@ctrerr\fi\relax}

\def\inst#1{\unskip$^{#1}$}
\def\fnmsep{\unskip$^,$}

\def\subtitle#1{\gdef\@subtitle{#1}}
\def\@subtitle{}

\def\headnote#1{\gdef\@headnote{#1}}
\def\@headnote{}

\def\thesaurus#1{\gdef\@thesaurus{#1}}
\def\@thesaurus{missing; you have not inserted them}%

\def\maketitle{\par
 \begingroup
 \def\thefootnote{\fnsymbol{footnote}}
 \if@twocolumn
   \twocolumn[\@maketitle]
 \else
   \newpage \@maketitle
 \fi
 \global\@topnum\z@
 \thispagestyle{empty}\@thanks
 \def\protect{\noexpand\protect\noexpand}
 \let\inst=\@gobble
 \let\fnmsep=\@gobble
 \let\thanks=\@gobble
 \xdef\@title{\@title\unskip}
 \def\stripauthor##1\and##2\endauthor{%
    \xdef\@author{##1\unskip\unskip\if!##2!\else\ et al.\fi}}
 \expandafter\stripauthor\@author\and\endauthor
 \endgroup
 \setcounter{footnote}{0}
 \setbox0=\hbox{\small\rm\@author\unskip: \@title\unskip}
 \dimen@=\wd0\relax
 \advance\dimen@ by 2cm
 \ifdim\dimen@>\textwidth
    \typeout{^^JAandA Warning: The running head built automatically from
             \string\author\space and \string\title
             ^^Jexceeds the pagewidth, please supply a shorter form
             ^^Jusing \string\markboth\string{...\string}\string{...\string}
             after the \string\maketitle-command.}%
    \setbox0=\hbox{\small\rm Please give a shorter version with:
             {\tt\string\markboth\string{...\string}\string{...\string}}}%
 \fi
 \setbox\runheadbox=\hbox{\unhbox0}
 \def\@oddhead{\small\rm\hfil\copy\runheadbox\hfil\llap{\thepage}}
 \def\@evenhead{\small\rm\rlap{\thepage}\hfil\copy\runheadbox\hfil}
 \let\maketitle\relax
 \let\@maketitle\relax
 \gdef\@thanks{}\gdef\@author{}\gdef\@title{}\gdef\@subtitle{}%
 \let\thanks\relax}

\def\AALogo{\setbox254=\hbox{ ASTROPHYSICS }%
\vbox{\baselineskip=10pt\hrule\hbox{\vrule\vbox{\kern3pt
\hbox to\wd254{\hfil ASTRONOMY\hfil}
\hbox to\wd254{\hfil AND\hfil}\copy254
\hbox to\wd254{\hfil\number\day.\number\month.\number\year\hfil}
\kern3pt}\vrule}\hrule}}
\def\makeheadbox{{\hbox to\textwidth{%
\hbox to0pt{\vbox{\hsize=30cc\baselineskip=12pt\hrule\hbox
to\hsize{\vrule\kern3pt\vbox{\kern3pt
\hbox{\bf A\&A manuscript no.}
\hbox{(will be inserted by hand later)}
\kern3pt\hrule\kern3pt\bf
\hbox{Your thesaurus codes are:}
\hbox{\rightskip=0pt plus3em\advance\hsize by-7pt
\vbox{\noindent\ignorespaces\@thesaurus}}
\kern3pt}\hfil\kern3pt\vrule}\hrule}\hss}
\hfil\llap{\quad\AALogo}}}}

\def\@maketitle{\newpage
 \rm\vbox to0pt{}\vskip-8mm
 \makeheadbox
\if!\@headnote!\else {\LARGE \it
  \vskip 8.89mm
  \pretolerance=10000
  \rightskip=0pt plus 3cm
  \noindent\@headnote \par}\vskip -3.7mm\fi
 \vskip11.712mm
 {\LARGE \bf\boldmath
  \pretolerance=10000
  \rightskip=0pt plus 4cm
  \noindent\ignorespaces
  \@title \par}\vskip .3cm
\if!\@subtitle!\else {\Large \bf\boldmath
  \vskip .05cm
  \pretolerance=10000
  \rightskip=0pt plus 3cm
  \noindent\@subtitle \par}\vskip10pt\fi
 {\bf \lineskip .5em
\setbox0=\vbox{\setcounter{@auth}{1}\def\and{\stepcounter{@auth}}%
\def\thanks##1{}\@author\global\c@@inst=\c@@auth}%
\def\lastand{\ifnum\c@@inst=2\relax\unskip{} and \else
\unskip, and \fi}%
\setcounter{@auth}{1}%
\def\and{\stepcounter{@auth}\ifnum\c@@auth=\c@@inst\lastand\else
\unskip, \fi}%
 \noindent\ignorespaces\@author\vskip.125cm}
 \small\rm
 \institutename
 \vskip .35cm
 \noindent\@date
 \par
 \vskip 21pt}

\mark{{}{}}

\def\markboth#1#2{%
\setbox0=\hbox{\small\rm\ignorespaces#1\unskip}
\dimen@=\wd0
\advance\dimen@ by 2cm\relax
\ifdim\dimen@>\textwidth
   \typeout{^^JAandA Warning: The running head you supplied in
            ^^Jusing \string\markboth\string{...\string}\string{...\string}
            ^^Jexceeds the pagewidth, please give a shorter form
            ^^Jafter the \string\maketitle-command.}%
   \setbox0=\hbox{\small\rm Please give a shorter version with:
            {\tt\string\markboth\string{...\string}\string{...\string}}}%
\fi
 \setbox\runheadbox=\hbox{\unhbox0}}

\def\ps@myheadings{\let\@mkboth=\@gobbletwo
\def\@oddhead{\small\rm\hfil\thepage\hfil
}\def\@oddfoot{}\def\@evenhead{\small\rm\hfil\thepage\hfil
}\def\@evenfoot{}\def\sectionmark##1{}\def\subsectionmark##1{}}
\newbox\runheadbox

\def\today{\ifcase\month\or
 January\or February\or March\or April\or May\or June\or
 July\or August\or September\or October\or November\or December\fi
 \space\number\day, \number\year}

\ps@myheadings 

\if@referee
  \onecolumn
\else
  \twocolumn
\fi
\def\@array[#1]#2{\setbox\@arstrutbox=\hbox{\vrule
     height\arraystretch \ht\strutbox
     depth\arraystretch \dp\strutbox
     width\z@}\@mkpream{@{}#2}\edef\@preamble{\halign \noexpand\@halignto
\bgroup \tabskip\z@ \@arstrut \@preamble \tabskip\z@ \cr}%
\let\@startpbox\@@startpbox \let\@endpbox\@@endpbox
  \if #1t\vtop \else \if#1b\vbox \else \vcenter \fi\fi
  \bgroup \let\par\relax
  \let\@sharp##\let\protect\relax \lineskip\z@\baselineskip\z@\@preamble}

\def\[{\relax\ifmmode\@badmath\else
 \begin{trivlist}%
 \@beginparpenalty\predisplaypenalty
 \@endparpenalty\postdisplaypenalty
 \item[]\leavevmode
 \hbox to\linewidth\bgroup $\m@th\displaystyle
 \hskip\mathindent\bgroup\fi}

\def\]{\relax\ifmmode \egroup $\hfil
       \egroup \end{trivlist}\else \@badmath \fi}

\def\equation{\@beginparpenalty\predisplaypenalty
  \@endparpenalty\postdisplaypenalty
\refstepcounter{equation}\trivlist \item[]\leavevmode
  \hbox to\linewidth\bgroup $\m@th
  \displaystyle
\hskip\mathindent}

\def\endequation{$\hfil
           \displaywidth\linewidth\@eqnnum\egroup \endtrivlist}

\def\eqnarray{\stepcounter{equation}\let\@currentlabel=\theequation
\global\@eqnswtrue
\global\@eqcnt\z@\tabskip\mathindent\let\\=\@eqncr
\abovedisplayskip\topsep\ifvmode\advance\abovedisplayskip\partopsep\fi
\belowdisplayskip\abovedisplayskip
\belowdisplayshortskip\abovedisplayskip
\abovedisplayshortskip\abovedisplayskip
$$\m@th\halign
to\linewidth\bgroup\@eqnsel\hskip\@centering$\displaystyle\tabskip\z@
  {##}$&\global\@eqcnt\@ne \hskip 2\arraycolsep \hfil${##}$\hfil
  &\global\@eqcnt\tw@ \hskip 2\arraycolsep $\displaystyle{##}$\hfil
   \tabskip\@centering&\llap{##}\tabskip\z@\cr}

\def\endeqnarray{\@@eqncr\egroup
      \global\advance\c@equation\m@ne$$\global\@ignoretrue
      }

\newdimen\mathindent
\def\abstract{\trivlist\item[\hskip\labelsep
{\bf Abstract.}]}

\def\keywords{\par\vspace{12pt}\noindent{\bf Key words: }}

\def\@cite#1#2{{#1\if@tempswa , #2\fi}}
\def\@biblabel#1{}

\def\squareforqed{\hbox{\rlap{$\sqcap$}$\sqcup$}}
\def\sq{\ifmmode\squareforqed\else{\unskip\nobreak\hfil
\penalty50\hskip1em\null\nobreak\hfil\squareforqed
\parfillskip=0pt\finalhyphendemerits=0\endgraf}\fi}

\def\la{\mathrel{\mathchoice {\vcenter{\offinterlineskip\halign{\hfil
$\displaystyle##$\hfil\cr<\cr\sim\cr}}}
{\vcenter{\offinterlineskip\halign{\hfil$\textstyle##$\hfil\cr
<\cr\sim\cr}}}
{\vcenter{\offinterlineskip\halign{\hfil$\scriptstyle##$\hfil\cr
<\cr\sim\cr}}}
{\vcenter{\offinterlineskip\halign{\hfil$\scriptscriptstyle##$\hfil\cr
<\cr\sim\cr}}}}}
\def\ga{\mathrel{\mathchoice {\vcenter{\offinterlineskip\halign{\hfil
$\displaystyle##$\hfil\cr>\cr\sim\cr}}}
{\vcenter{\offinterlineskip\halign{\hfil$\textstyle##$\hfil\cr
>\cr\sim\cr}}}
{\vcenter{\offinterlineskip\halign{\hfil$\scriptstyle##$\hfil\cr
>\cr\sim\cr}}}
{\vcenter{\offinterlineskip\halign{\hfil$\scriptscriptstyle##$\hfil\cr
>\cr\sim\cr}}}}}

\def\arcsec{\hbox{$^{\prime\prime}$}}
\def\utw{\smash{\rlap{\lower5pt\hbox{$\sim$}}}}
\def\udtw{\smash{\rlap{\lower6pt\hbox{$\approx$}}}}

\def\farcs{\hbox{$.\!\!^{\prime\prime}$}}

\def\diameter{{\ifmmode\mathchoice
{\ooalign{\hfil\hbox{$\displaystyle/$}\hfil\crcr
{\hbox{$\displaystyle\mathchar"20D$}}}}
{\ooalign{\hfil\hbox{$\textstyle/$}\hfil\crcr
{\hbox{$\textstyle\mathchar"20D$}}}}
{\ooalign{\hfil\hbox{$\scriptstyle/$}\hfil\crcr
{\hbox{$\scriptstyle\mathchar"20D$}}}}
{\ooalign{\hfil\hbox{$\scriptscriptstyle/$}\hfil\crcr
{\hbox{$\scriptscriptstyle\mathchar"20D$}}}}
\else{\ooalign{\hfil/\hfil\crcr\mathhexbox20D}}%
\fi}}

\def\bbbc{{\mathchoice {\setbox0=\hbox{$\displaystyle\rm C$}\hbox{\hbox
to0pt{\kern0.4\wd0\vrule height0.9\ht0\hss}\box0}}
{\setbox0=\hbox{$\textstyle\rm C$}\hbox{\hbox
to0pt{\kern0.4\wd0\vrule height0.9\ht0\hss}\box0}}
{\setbox0=\hbox{$\scriptstyle\rm C$}\hbox{\hbox
to0pt{\kern0.4\wd0\vrule height0.9\ht0\hss}\box0}}
{\setbox0=\hbox{$\scriptscriptstyle\rm C$}\hbox{\hbox
to0pt{\kern0.4\wd0\vrule height0.9\ht0\hss}\box0}}}}
\def\bbbq{{\mathchoice {\setbox0=\hbox{$\displaystyle\rm
Q$}\hbox{\raise
0.15\ht0\hbox to0pt{\kern0.4\wd0\vrule height0.8\ht0\hss}\box0}}
{\setbox0=\hbox{$\textstyle\rm Q$}\hbox{\raise
0.15\ht0\hbox to0pt{\kern0.4\wd0\vrule height0.8\ht0\hss}\box0}}
{\setbox0=\hbox{$\scriptstyle\rm Q$}\hbox{\raise
0.15\ht0\hbox to0pt{\kern0.4\wd0\vrule height0.7\ht0\hss}\box0}}
{\setbox0=\hbox{$\scriptscriptstyle\rm Q$}\hbox{\raise
0.15\ht0\hbox to0pt{\kern0.4\wd0\vrule height0.7\ht0\hss}\box0}}}}
\def\bbbt{{\mathchoice {\setbox0=\hbox{$\displaystyle\rm
T$}\hbox{\hbox to0pt{\kern0.3\wd0\vrule height0.9\ht0\hss}\box0}}
{\setbox0=\hbox{$\textstyle\rm T$}\hbox{\hbox
to0pt{\kern0.3\wd0\vrule height0.9\ht0\hss}\box0}}
{\setbox0=\hbox{$\scriptstyle\rm T$}\hbox{\hbox
to0pt{\kern0.3\wd0\vrule height0.9\ht0\hss}\box0}}
{\setbox0=\hbox{$\scriptscriptstyle\rm T$}\hbox{\hbox
to0pt{\kern0.3\wd0\vrule height0.9\ht0\hss}\box0}}}}
\def\bbbs{{\mathchoice
{\setbox0=\hbox{$\displaystyle     \rm S$}\hbox{\raise0.5\ht0\hbox
to0pt{\kern0.35\wd0\vrule height0.45\ht0\hss}\hbox
to0pt{\kern0.55\wd0\vrule height0.5\ht0\hss}\box0}}
{\setbox0=\hbox{$\textstyle        \rm S$}\hbox{\raise0.5\ht0\hbox
to0pt{\kern0.35\wd0\vrule height0.45\ht0\hss}\hbox
to0pt{\kern0.55\wd0\vrule height0.5\ht0\hss}\box0}}
{\setbox0=\hbox{$\scriptstyle      \rm S$}\hbox{\raise0.5\ht0\hbox
to0pt{\kern0.35\wd0\vrule height0.45\ht0\hss}\raise0.05\ht0\hbox
to0pt{\kern0.5\wd0\vrule height0.45\ht0\hss}\box0}}
{\setbox0=\hbox{$\scriptscriptstyle\rm S$}\hbox{\raise0.5\ht0\hbox
to0pt{\kern0.4\wd0\vrule height0.45\ht0\hss}\raise0.05\ht0\hbox
to0pt{\kern0.55\wd0\vrule height0.45\ht0\hss}\box0}}}}
\def\bbbz{{\mathchoice {\hbox{$\sf\textstyle Z\kern-0.4em Z$}}
{\hbox{$\sf\textstyle Z\kern-0.4em Z$}}
{\hbox{$\sf\scriptstyle Z\kern-0.3em Z$}}
{\hbox{$\sf\scriptscriptstyle Z\kern-0.2em Z$}}}}

\def\typeset{\vfill{\small\rm\noindent This article was processed by the
                    author using Springer-Verlag \LaTeX\ A\&A style file
                    {\it L-AA\/} version 3.\par}}

\def\enddocument{\par\typeset
                 \@checkend{document}\clearpage\begingroup
\if@filesw \immediate\closeout\@mainaux
\def\global\@namedef##1##2{}\def\newlabel{\@testdef r}%
\def\@testbibciteaa ##1##2##3{\def\@tempa{##3}%
\if!##2!\else
 \expandafter\ifx \csname##1@##2\endcsname \@tempa \else \@tempswatrue\fi
\fi}\def\bibcite{\@testbibciteaa b}%
\@tempswafalse\makeatletter\input \jobname.aux
\if@tempswa \@warning{Label(s) may have changed. Rerun to get
cross-references right}\fi\fi\endgroup\deadcycles\z@\@@end}
\endinput

\documentstyle[bibstyle]{l-aa}

\newcommand{\beq}{\begin{equation}}
\newcommand{\eeq}{\end{equation}}

\newcommand{\tabref}[1]{Table~\ref{#1}}
\newcommand{\figref}[1]{Fig.~\ref{#1}}

\newcommand{\eqref}[1]{Eq.(\ref{#1})}

\newcommand{\secref}[1]{Sect.~\ref{#1}}

\newcommand{\Tabref}[1]{\tabref{#1}}
\newcommand{\Figref}[1]{Figure~\ref{#1}}

\newcommand{\Eqref}[1]{Equation~(\ref{#1})}

\newcommand{\about}{\mbox{$\sim$\,}}	

\def\msol{\mbox{M$_\odot$}}
\def\deg{\hbox{$^\circ$}}

\def\la{\mathrel{\hbox{\rlap{\hbox{\lower4pt\hbox{$\sim$}}}\hbox{$<$}}}}
\def\ga{\mathrel{\hbox{\rlap{\hbox{\lower4pt\hbox{$\sim$}}}\hbox{$>$}}}}
\def\sq{\hbox{\rlap{$\sqcap$}$\sqcup$}}

\def\arcsec{\hbox{$^{\prime\prime}$}}

\def\farcs{\hbox{$.\!\!^{\prime\prime}$}}

\newcommand{\down}[2]{\mbox{$#1_{\mbox{\scriptsize #2}}$}}

\newcommand{\mc}[3]{\multicolumn{#1}{#2}{#3}}
\newcommand{\z}{\phantom{0}}
\newcommand{\ra}{\rlap{$^a$}}

\newcommand{\SGHz}{\down{S}{5GHz}}

\newcommand{\Mion}{\down{M}{ion}}

\newcommand{\dopt}{\down{\theta}{opt}}
\newcommand{\drad}{\down{\theta}{rad}}
\newcommand{\dpsf}{\down{\theta}{PSF}}
\newcommand{\dgauss}{\down{\theta}{gauss}}
\newcommand{\dfwhm}{\down{\theta}{FWHM}}
\newcommand{\dtrue}{\down{\theta}{true}}
\newcommand{\dsphere}{\down{\theta}{sphere}}
\newcommand{\ddisk}{\down{\theta}{disk}}
\newcommand{\dshell}{\down{\theta}{shell}}
\newcommand{\dten}{\down{\theta}{10\%}}

\input{psfig}

\begin{document}

\thesaurus{8          
           (09.16.1)} 

\title{Angular diameters of compact planetary nebulae}

\author{Timothy R. Bedding \and Albert A. Zijlstra}

\institute{
European Southern Observatory, Karl-Schwarzschild-Str.~2,
D-85748 Garching bei M\"unchen, Federal Republic of Germany}

\date{Received................; Accepted.................}

\offprints{T.R. Bedding}

\maketitle

\markboth{T.R. Bedding \&\ A.A. Zijlstra: Angular diameters of planetary
nebulae}{}

\begin{abstract}
We have obtained H$\alpha$ CCD images of 21 compact planetary nebulae in
the galactic bulge.  All objects are resolved and, after deconvolving the
seeing disk, we find diameters of 1--2$^{\prime\prime}$ with typical
uncertainties of 20\%.  The values are generally in good agreement with
radio measurements but are smaller than older optical measurements.  Based
on the new values, we find that the Shklovsky method for distance
determination greatly overestimates the actual distance for all of our
objects.


\end{abstract}

\section{Introduction}

The diameter of a planetary nebula is an important parameter.  It is needed
for modelling the optical spectrum and for studying the evolution of the
nebula and its central star.  From the diameter and the H$\beta$ flux of
the nebula, its density and mass can be calculated.  Alternatively, if a
value for the mass is assumed, the distance can be derived from the
measured angular diameter.  This is the basis of the so-called Shklovsky
method of distance determination, which assumes that all planetary nebulae
(PN) have the same mass (\about0.2~M$_\odot$).  However, such statistical
methods are thought to be inaccurate.  Many studies have therefore
concentrated on objects in the galactic bulge or the Magellanic clouds,
where the distance is known.  PN in the Magellanic clouds are mostly faint
and small ($<$1$''$) and diameters are only available for a limited sample
(\citebare{WBD86}, \citeyear{WMD87}).  Galactic-bulge nebulae are somewhat
easier to measure and many more data have been published.

Angular diameter determinations have been hampered by several problems.  At
the distance of the bulge, many PN are only a few arcseconds across and
therefore require careful measurements.  Also, PN show a wealth of
different morphologies, making it difficult to assign unambiguous
diameters.  Several different definitions appear in the literature, such as
measuring the outermost position at which emission is seen, the 10\% (or
25\%) surface-brightness contour, or the FWHM\@.  Finally, the ionisation
structure makes the diameter dependent on the particular emission line
observed.

\citeone[hereafter STAS91]{STA91} have noticed that large discrepancies
exist between optical and radio angular diameters.  This is especially true
for compact objects, where radio diameters are systematically smaller.
{}From optical surveys it appears that few bulge PN are smaller than $4''$,
while radio diameters are sometimes as small as $1''$.  STAS91 conclude
that the radio diameters are less reliable because they are derived with
model-dependent deconvolution techniques.  This conclusion is disputed by
\citeone[hereafter PZ92]{P+Z92}, who argue that optical determinations
often show large intrinsic uncertainties, especially for small objects.
They reason that the optical diameters have mainly been measured from
photographic and even objective-prism plates, using procedures that are
unreliable for small nebulae.

Clearly this discrepancy has important consequences for the physical
characteristics and inferred evolution of bulge PN\@.  In an attempt to
resolve this question, we have obtained narrow-band H$\alpha$ CCD images of
a sample of bulge PN, concentrating on objects for which the discrepancy
between catalogued optical and radio diameters is large.  Before describing
these observations, we briefly examine the extent of the disagreement
between optical and radio measurements and show that it is even greater
than indicated by STAS91\@.

\section{A comparison of optical and radio diameters}
\label{sec.optical-radio}

Many radio diameters (\drad) of PN have been obtained from VLA imaging
surveys \cite{ZPB89,A+K90,R+P91}, while optical diameters (\dopt) have
generally been measured from photographic plates (including objective-prism
plates).  A compilation of available diameter measurements can be found in
the Strasbourg--ESO Catalogue \cite{AOS92}.  Note that the errata to this
catalogue lists many objects for which printed values of \dopt\ are
actually only upper limits.

\begin{figure*}
\centerline{
\psfig{bbllx=10mm,bblly=57mm,bburx=202mm,bbury=244mm,%
figure=dummy,height=\the\hsize,scale=32}
\hfill
\psfig{bbllx=10mm,bblly=57mm,bburx=202mm,bbury=244mm,%
figure=dummy,height=\the\hsize,scale=32}
\hfill
\psfig{bbllx=8mm,bblly=57mm,bburx=206mm,bbury=244mm,%
figure=dummy,height=\the\hsize,scale=32}
}
\caption[]{
Comparison of catalogued radio and optical diameters (in arcsec) of
galactic PN.  Optical upper limits are for objects denoted ``stellar'' and
triangles show PN observed in our program.

\label{fig.diams}}

\end{figure*}

In \figref{fig.diams} we compare optical and radio diameters based on
values in the Strasbourg--ESO Catalogue.  \Figref{fig.diams}a is for
galactic-bulge PN: this plot is similar to Fig.~2 of STAS91, in which they
show 52 bulge objects for which optical and radio diameters are available
(where STAS91 plot radii rather than diameters).  Using the same
criteria for galactic-bulge membership as STAS91 and PZ92 ($|l|\!  <
\!10\deg$, $|b|\! < \!10\deg$, 6\,cm radio flux density $<\!100$\,mJy and
angular diameter $<\!20''$), we find 73 galactic-bulge PN for which both
\dopt\ and \drad\ are given.  We do not show objects for which only upper
limits or uncertain values are listed.  However, \figref{fig.diams}a
does include 17~PN which have radio diameters but are listed as being
``stellar'' in the optical.  They are plotted as upper limits using the
arbitrary value of $\dopt=1.5''$.

The triangles in \figref{fig.diams} show the objects observed in our
program.  Some of these have optical diameters which are uncertain, so they
would not otherwise appear in these figures.  We include them to show the
properties of our sample.  Two program objects are not shown: one only has
an upper limit for \dopt\ and the other has no measured radio diameter.

A fuller comparision of optical and radio measurements is possible if we do
not require galactic-bulge membership.  Membership of the bulge is not
relevant to the present discussion, since our aim is to compare the
accuracy of optical and radio measurements.  There are 155 objects that do
not satisfy the above criteria for bulge membership but which nevertheless
have catalogued values for \dopt\ and \drad, plus a further 36 that are
``stellar'' in the optical.  \Figref{fig.diams}b shows the smallest of
these objects, while \figref{fig.diams}c shows larger (and probably nearby)
PN on a logarithmic scale.

For the larger PN, we expect the optical measurement to be the most precise,
since the radio observations of these objects were made at lower
resolution.  The uncertainties in \drad\ should be a fixed percentage of
the measured value, regardless of the size.  On a logarithmic scale, this
corresponds to error bars of constant length, which is the reason for
showing these data with logarithmic axes.  Bearing this in mind, we see
that the two methods are in reasonable agreement and that there is no trend
for one method to produce systematically larger values than the other.  The
20--30\% scatter between \dopt\ and \drad\ is also partly caused by the
different conventions adopted by each observer when defining the diameter.

Turning to the smaller objects (\dopt$<$10\arcsec), we see that the
agreement is often very poor and that optical diameters are systematically
larger than radio ones.  If the radio measurements are more reliable, we
would expect optical diameters measured using modern detectors to be in
much better agreement with radio measurements than are those made using
photographic plates.  Support for this view is given by \citeone{DHT90},
who imaged six compact PN with a photon-counting detector using
shift-and-add techniques.  They did find very small sizes and their
diameters were in agreement with radio determinations for all but one
object, which is larger in the radio image.  However these authors observed
the [OIII] line, which does not always trace the full nebular extent.  Many
PN appear smaller in [OIII] and HeII, and larger in [NII], depending on the
evolutionary phase of the nebula.  The morphologies can also be different
in different lines, as shown in CCD images obtained by \citeone{Bal87}.
Since hydrogen is by far the main constituent of PN, it follows that
H$\alpha$ is the best tracer of the mass distribution.  This is especially
relevant when one uses the angular diameter for calculating masses or
Shklovsky distances, as discussed in \secref{sec.Shklovsky} below.  We have
therefore chosen to observe using a narrow H$\alpha$ filter that excludes
the [NII] lines.

\begin{figure*}

\centerline{
\hfill
\psfig{width=\the\hsize,clip=,scale=30,%
bbllx=37mm,bblly=147mm,bburx=163mm,bbury=272mm,%
figure=dummy}
\hfill
\psfig{width=\the\hsize,clip=,scale=30,%
bbllx=37mm,bblly=147mm,bburx=163mm,bbury=272mm,%
figure=dummy}
\hfill
\psfig{width=\the\hsize,clip=,scale=30,%
bbllx=37mm,bblly=147mm,bburx=163mm,bbury=272mm,%
figure=dummy}
\hfill
}

\centerline{
\hfill
\psfig{width=\the\hsize,clip=,scale=30,%
bbllx=37mm,bblly=147mm,bburx=163mm,bbury=272mm,%
figure=dummy}
\hfill
\psfig{width=\the\hsize,clip=,scale=30,%
bbllx=37mm,bblly=147mm,bburx=163mm,bbury=272mm,%
figure=dummy}
\hfill
\psfig{width=\the\hsize,clip=,scale=30,%
bbllx=37mm,bblly=147mm,bburx=163mm,bbury=272mm,%
figure=dummy}
\hfill
}

\centerline{
\hfill
\psfig{width=\the\hsize,clip=,scale=30,%
bbllx=37mm,bblly=147mm,bburx=163mm,bbury=272mm,%
figure=dummy}
\hfill
\psfig{width=\the\hsize,clip=,scale=30,%
bbllx=37mm,bblly=147mm,bburx=163mm,bbury=272mm,%
figure=dummy}
\hfill
\psfig{width=\the\hsize,clip=,scale=30,%
bbllx=37mm,bblly=147mm,bburx=163mm,bbury=272mm,%
figure=dummy}
\hfill
}

\caption[]{
H$\alpha$ images with contour levels at 1, 2, 5, 10, 20, 35, 50, 65 and
80\% of the nebula peak.  All images are $20''\times20''$ with north up and
east to the left.  To filter the noise, the images have been smoothed with
a 5-pixel (0\farcs65) box prior to plotting.  For comparison, the inset in
the first figure shows a gaussian with a FWHM of 1.2$''$, displayed using
the same scale and contour levels as the PN images.
\label{fig.contour1}}

\end{figure*}

\begin{figure*}

\centerline{
\hfill
\psfig{width=\the\hsize,clip=,scale=30,%
bbllx=37mm,bblly=147mm,bburx=163mm,bbury=272mm,%
figure=dummy}
\hfill
\psfig{width=\the\hsize,clip=,scale=30,%
bbllx=37mm,bblly=147mm,bburx=163mm,bbury=272mm,%
figure=dummy}
\hfill
\psfig{width=\the\hsize,clip=,scale=30,%
bbllx=37mm,bblly=147mm,bburx=163mm,bbury=272mm,%
figure=dummy}
\hfill
}

\centerline{
\hfill
\psfig{width=\the\hsize,clip=,scale=30,%
bbllx=37mm,bblly=147mm,bburx=163mm,bbury=272mm,%
figure=dummy}
\hfill
\psfig{width=\the\hsize,clip=,scale=30,%
bbllx=37mm,bblly=147mm,bburx=163mm,bbury=272mm,%
figure=dummy}
\hfill
\psfig{width=\the\hsize,clip=,scale=30,%
bbllx=37mm,bblly=147mm,bburx=163mm,bbury=272mm,%
figure=dummy}
\hfill
}

\centerline{
\hfill
\psfig{width=\the\hsize,clip=,scale=30,%
bbllx=37mm,bblly=147mm,bburx=163mm,bbury=272mm,%
figure=dummy}
\hfill
\psfig{width=\the\hsize,clip=,scale=30,%
bbllx=37mm,bblly=147mm,bburx=163mm,bbury=272mm,%
figure=dummy}
\hfill
\psfig{width=\the\hsize,clip=,scale=30,%
bbllx=37mm,bblly=147mm,bburx=163mm,bbury=272mm,%
figure=dummy}
\hfill
}

\centerline{
\hfill
\psfig{width=\the\hsize,clip=,scale=30,%
bbllx=37mm,bblly=147mm,bburx=163mm,bbury=272mm,%
figure=dummy}
\hfill
\psfig{width=\the\hsize,clip=,scale=30,%
bbllx=37mm,bblly=147mm,bburx=163mm,bbury=272mm,%
figure=dummy}
\hfill
\psfig{width=\the\hsize,clip=,scale=30,%
bbllx=37mm,bblly=147mm,bburx=163mm,bbury=272mm,%
figure=dummy}
\hfill
}

\addtocounter{figure}{-1}
\caption[]{{\bf (cont.)}
For some images having lower signal-to-noise (marked {\tt *}), the two
lowest contours are omitted and the smoothing box is 6~pixels instead of 5.
}

\end{figure*}

\section{Observations and data processing}

We observed 21 galactic PN during 3.5~hours on 1993 May 11, using the ESO
3.5-m New Technology Telescope (NTT) in Chile.  The observations were made
from ESO Headquarters in Garching using the NTT remote-control system
\cite{BBC93}.  We used the SUSI camera (SUperb Seeing Imager), which has a
1024$^2$-pixel Tektronix CCD and an image scale of 0\farcs13/pixel, in
combination with an H$\alpha$ filter centred at 6561\AA\ with a FWHM of
12\AA\@.

To reduce the CCD readout time and hence improve the observing duty cycle,
we only read out the central $400\times400$ pixels ($52''\times52''$) of
the detector.  This also increased the speed at which data frames could be
transferred to Garching, allowing us to inspect each image shortly after it
was taken.  Calibration of the CCD images followed standard procedures:
subtraction of the median of several bias frames, division by the median of
several dome flat-field exposures and subtraction of the background sky
level.  In \figref{fig.contour1} we show $20''\times20''$ images of each
object, with contours expressed as percentages of the peak nebula flux.
Problems with telescope pointing caused some of our objects to appear near
the edge of the CCD window; in one case, the object was truncated.

\begin{table*}
\caption[]{Angular diameters of 21 galactic planetary nebulae\label{tab.obs}}
\begin{flushleft}
\begin{tabular}{
l                             c                    c        l
   c     c        c      c             c         c   c      l             l
  }
    \hline\noalign{\smallskip}
Object                  & \dpsf           & \mc{2}{c}{\dgauss}
 &   &\dsphere &\ddisk &\dshell&\dten          &   &\drad &  \dopt     & Notes
  \\
           \noalign{\smallskip}
    \hline\noalign{\smallskip}
321.0$+$03.9\rlap{$^a$} & $1.3\z\pm0.3\z$ &
$1.6$&$\!\!\!\!\!\!\pm\,0.4$ &   & 2.7     & 2.4   & 2.3   &               &
& ---  &       st.  &  \\                
350.9$+$04.4            & $1.17 \pm0.04 $ &
$1.7$&$\!\!\!\!\!\!\pm\,0.1$ &   & 2.9     & 2.5   & 2.4   &               &
& 2.2  &       5.6  &          \\        
351.1$+$04.8            & $1.2\z\pm0.3\z$
&$2.1\times1.9$&$\!\!\!\!\!\!\pm\,0.3$ &   & 3.4     & 2.9   & 2.7   &
     &   & 2.6  &       8:   &          \\        
353.3$+$06.3            & $1.25 \pm0.04 $
&$1.7\times1.3$&$\!\!\!\!\!\!\pm\,0.1$ &   & 2.4     & 2.1   & 2.0   &
     &   & 1.6  &       8:   &          \\        
355.4$-$02.4            & $1.40 \pm0.04 $ &              &
 &   &         &       &       &$5.6\times4.8$ &   & 2.8  &       7.2  &bipolar
\\          
355.7$-$03.5            & $1.3\z\pm0.1\z$ &
$1.3$&$\!\!\!\!\!\!\pm\,0.2$ &   & 2.3     & 2.0   & 1.9   &               &
& 1.1  &       2.0  &          \\        
355.9$+$03.6            & $1.20 \pm0.03 $ &
$0.6$&$\!\!\!\!\!\!\pm\,0.1$ &   & 1.1     & 0.9   & 0.9   &               &
& 0.7  &       7:   &          \\        
356.2$-$04.4            & $1.4\z\pm0.1\z$ &
$1.5$&$\!\!\!\!\!\!\pm\,0.2$ &   & 2.6     & 2.3   & 2.2   &               &
& 1.7  &       2.4  &          \\        
356.9$+$04.5            & $1.10 \pm0.05 $ &
$1.1$&$\!\!\!\!\!\!\pm\,0.1$ &   & 1.9     & 1.7   & 1.6   &               &
& 2.7  &       15:  &core + bipolar\\    
357.2$+$07.4            & $1.15 \pm0.04 $ &
$1.1$&$\!\!\!\!\!\!\pm\,0.1$ &   & 1.9     & 1.7   & 1.6   &               &
& 1.4  &       8:   &          \\        
357.4$-$03.2\rlap{$^b$} & $1.1\z\pm0.1\z$ &
$1.8$&$\!\!\!\!\!\!\pm\,0.1$ &   & 3.0     & 2.6   & 2.5   &               &
& 2.2  &       5.4  &          \\        
358.2$+$04.2            & $1.24 \pm0.07 $ &
$2.3$&$\!\!\!\!\!\!\pm\,0.3$ &   & 3.9     & 3.4   & 3.1   &               &
& 2.9  &       5.4  &\\                  
358.8$+$04.0\rlap{$^c$} & $1.09 \pm0.05 $ &              &
 &   &         &       &       &$7.5\times7.0$ &   & 6.0  &       st.
&peanut-shaped\\     
359.8$+$03.7            & $1.11 \pm0.04 $ &
$1.4$&$\!\!\!\!\!\!\pm\,0.1$ &   & 2.4     & 2.1   & 2.0   &               &
& 1.8  &       7:   &          \\        
000.1$+$17.2            & $1.2\z\pm0.4\z$
&$1.9\times1.6$&$\!\!\!\!\!\!\pm\,0.5$ &   & 2.7     & 2.4   & 2.3   &
     &   & 1.8  &       4.6  &  \\                
000.2$-$01.9            & $1.76 \pm0.05 $ &              &
 &   &         &       &       &$11\times9$    &   & 5.0  &       8.2  &
elliptical + halo\\
003.1$+$03.4            & $1.10 \pm0.03 $ &              &
 &   &         &       &       &4.0            &   & 4.0  &       3.6  &
  \\        
004.6$+$06.0            & $1.26 \pm0.04 $ &              &
 &   &         &       &       &$5.5\times4.8$ &   & 4.7  &       8.6  &
  \\        
005.0$+$04.4            & $1.30 \pm0.04 $ &
$0.7$&$\!\!\!\!\!\!\pm\,0.1$ &   & 1.3     & 1.1   & 1.1   &               &
& 0.8  &       5.2  & image trailed\\    
006.4$+$02.0            & $1.8\z\pm0.1\z$ &
$1.4$&$\!\!\!\!\!\!\pm\,0.4$ &   & 2.5     & 2.2   & 2.1   &               &
& 7.0  &       st.  &\drad\ unreliable \\
008.2$+$06.8            & $1.17 \pm0.03 $ &
$0.7$&$\!\!\!\!\!\!\pm\,0.1$ &   & 1.3     & 1.1   & 1.1   &               &
& 1.0  &\llap{$<$}10& image trailed\\    
\noalign{\smallskip}
\hline
\end{tabular}\\
All exposure times 100\,s except: $^a$30\,s, $^b$20\,s, $^c$300\,s.
Diameters are in arcseconds and are defined in \secref{sec.diams}.
\end{flushleft}
\end{table*}

\begin{figure}
\centerline{
\psfig{bbllx=10mm,bblly=57mm,bburx=202mm,bbury=244mm,%
figure=dummy,width=\the\hsize,scale=95}
}
\caption[]{

Relationship between the true diameter (\dtrue) and the gaussian-fitted
FWHM (\dgauss) for test objects after convolving with a point-spread
function.  The points are data from a computer model (see text).  For
comparison, the dashed line shows the relation $\dtrue/\dgauss=1.6$.

\label{fig.fudge}}

\end{figure}

All objects are resolved and in some cases, the morphology of the nebula is
apparent.  Data for two objects (005.0$+$04.4 and 008.2$+$06.8) were taken
with the auto\-guider disactivated and the resulting images show an
elongation of both the nebulae and field stars which is not real.  For the
remaining objects, the structures seen in our images (elliptical, bipolar
and peanut-shaped) are genuine; they can be compared with the evolutionary
sequences of \citeone{Bal87}.

\section{Diameter determinations} \label{sec.diams}

The images presented here are substantially broadened by the seeing and, to
some extent, by guiding and focussing errors.  This must be taken into
account when deriving angular diameters and we now describe the method we
have used to deconvolve the point-spread function.

The point-spread function is well approximated by a gaussian with a FWHM of
\dpsf, and \tabref{tab.obs} lists this parameter for each image.  These
values, determined by fitting gaussian profiles to field stars, are mostly
in the range 1.1--1.4$''$.  We can also measure \dfwhm, the FWHM of the
nebula image itself.  If the object really were gaussian, the deconvolution
would be trivial: the true FWHM of the nebula would simply be
 \beq
    \dgauss = \sqrt{(\dfwhm)^2 - (\dpsf)^2}. \label{eq.dgauss}
 \eeq
 However, a gaussian profile is generally a poor description of a planetary
nebula and we will do better to adopt a more realistic model of the density
distribution.  In the case of a partially-resolved object, the simplest
approach is to calculate \dgauss\ and then apply a correction factor based
on an assumed density distribution.  This method is discussed by
\citeone{P+W78} in the context of radio-source angular diameters.

We have calculated the correction factor for three different models, each
of which has a well-defined diameter.  The models are: a sphere of uniform
density and diameter \dsphere, a disk of uniform surface brightness and
diameter \ddisk, and a uniform-density shell with fractional thickness 0.2
and outer diameter \dshell\ (i.e., inner diameter 0.8\,\dshell).  The shell
model repesents a nebula with a volume filling factor of 0.5.

The calculations were done by making artificial images and convolving them
with a gaussian point-spread function.  We then measured the FWHM of the
resulting images, using the same gaussian-fitting program as used for the
real data.  The results for the three models are plotted in
\figref{fig.fudge}.  Here, \dtrue\ is the diameter of the original model
(either \dsphere, \ddisk\ or \dshell) and \dgauss\ is calculated from the
measured FWHM using \eqref{eq.dgauss}.

The correction factor to be applied for the particular choice of model is
$\dtrue/\dgauss$.  This factor varies from \about1.8 for barely-resolved
objects to \about1.4 for a well-resolved shell.  However, the deconvolution
procedure described here should not be applied to well-resolved sources.
Their diameters should be measured directly from the image.

Although the models we have examined are more realistic than a gaussian, PN
are still expected to have morphologies that are considerably more
complicated.  However, the range of correction factors for these models
gives a good indication of the uncertainties in the derived diameters.  For
example, in the case where \dgauss\ is equal to \dpsf, the correction
factors show a range of about 20\% between the three models.

\subsection{Results}

\Tabref{tab.obs} lists the diameters we derive for all three models.  A few
objects are sufficiently resolved that deconvolution is not necessary.
These objects are listed with a 10\%-contour diameter in Table 1, which is
the diameter finally adopted.  For other sources, the model diameters show
a range of about 20\% for each object.

Columns 8 and 9 of \tabref{tab.obs} list the catalogued radio and optical
diameters.  Our derived diameters agree reasonably well with the radio
values, but are substantially smaller than previous optical measurements.
One object (006.4$+$02.0) is discrepant.  The catalogued radio diameter of
$7''$ originates from \citeone{ZPB89} and, after inspecting the original
radio data, we found that it was observed at a resolution of only $8''$ and
that the listed radio diameter is actually an upper limit.  Most of the
other sources were observed in the radio at a resolution of \about2$''$.
The resolution of our H$\alpha$ images is significantly better and the new
diameters should be more accurate than the radio values.

Emission-line images of a large number of PN have been published by
\citeone{SCM92} and three of these objects are also in our sample
(350.9+04.4, 000.1+17.2 and 003.1+03.4).  Their observations were made in
[OIII] and H$\alpha$, although the filter used for the latter images also
included the [NII] lines.  They list angular diameters for these objects,
but the values were generally derived by estimating the maximum extent of
the emission (Corradi, private communication).  While this may be adequate
for extended objects, the diameters they obtain for compact PN are probably
not good estimates.  For the three objects we have in common, these authors
have kindly made their images available to us and we have analysed them in
the way described above, by fitting Gaussian profiles and deconvolving the
seeing disk.  The diameters we obtain from these H$\alpha$/[NII] images are
similar to, but slightly larger than, those in \tabref{tab.obs}.  This is
possibly due to the presence of [NII] emission.

To summarize, we find that most of our H$\alpha$ diameters are in good
agreement with the radio measurements.  \Figref{fig.diams-ours} shows the
comparison.  The optical diameters in this plot are those derived from the
shell model, except that we have used the 10\%-contour diameter for five
well-resolved objects.  The shell model is arguably the most realistic of
the three PN models and, indeed, the higher resolution images of
\citeone{DHT90} tend to support this.  In addition, inspecting the results
in \tabref{tab.obs} suggests that this model generally gives the best
agreement with the radio diameters.

\begin{figure}
\centerline{
\psfig{bbllx=8mm,bblly=57mm,bburx=206mm,bbury=244mm,%
figure=dummy,width=\the\hsize,scale=95}
}
\caption[]{
Comparison of the diameters derived in this paper with catalogued radio
values.
\label{fig.diams-ours}}
\end{figure}

\section{Shklovsky distances and mass determinations}\label{sec.Shklovsky}

\subsection{A review of the method}

\citeone{Shk56} proposed a simple statistical method to derive the distance
to planetary nebulae.  He assumed that all PN can be approximated by
spheres of fully ionized hydrogen with the same mass $\Mion$, a uniform
density (which varies between nebulae) and a filling factor $\varepsilon$.
The total ionized mass of the nebula can be shown to be
 \beq
   \Mion = 2.60\times10^{-5} (\SGHz)^{0.5} \varepsilon^{0.5}
         \theta^{-1.5} d^{2.5}\quad \msol, \label{eq.Shklovsky}
 \eeq
 where $\SGHz$ is the radio flux in mJy, $d$ is the distance in kpc and
$\theta$ is the angular diameter in arcsec.  This assumes an electron
temperature of $10^4$\,K and a helium/hydrogen number-density ratio
$n$(He)/$n$(H)$=0.10$.

\Eqref{eq.Shklovsky} shows that the derived distance is not a very strong
function of either the assumed mass or the measured diameter.  For this
reason, it is often assumed that the method yields reasonable distances,
even though the underlying assumptions are not fully correct.  There are
two main assumptions: firstly, that the event leading to the formation of
the circumstellar shell always ejects the same amount of mass,
independently of stellar parameters; and secondly, that the circumstellar
shell is always fully ionized.  Clearly, these assumptions are
simplifications.  The amount of mass lost by the star must be a strong
function of stellar mass, since the central stars of PN show a very limited
range of masses (0.5--0.8\,\msol), while the progenitor masses range from
1--8\,\msol.  Also, emission from molecules (H$_2$, CO, OH) has been detected
from a number of PN which therefore are not fully ionized \cite{Rod87}.
However, if most PN are reasonably alike, the Shklovsky method would
still give useful distances.

Another uncertainty is introduced by the choice of the constants $\Mion$
and $\varepsilon$.  Values for the nebular mass in the recent literature
range from 0.16 to 0.40\,\msol, depending on which objects and which
techniques are used to calibrate the distances (e.g., \citebare{Wei89}).
The filling factor $\varepsilon$ is normally taken as approximately 0.5,
based on comparisons between densities derived from forbidden-line ratios
and from continuum measurements (\citebare{Pot92}, \citeyear{Pot84}). The
mass and filling factor, however, are not independent: the value assumed
for each depends on the other.

In view of the uncertain assumptions behind the Shklovsky method,
\citeone{Dau82} has proposed a distance scale which assumes that PN only
become fully ionized when the radius reaches 0.10~pc: nebulae smaller than
this will be ionization bounded and have an ionized mass lower than the
Shklovsky mass; larger nebulae are density bounded and their ionized mass
will be equal to the total (Shklovsky) mass.  This view has since been
taken by many authors (e.g., \citebare{Kwo86}; \citebare{Sta89};
\citebare{CKS92}).  However, \citename{Kwo86} has shown that this method
cannot produce unambiguous distances unless further information on the
evolutionary status is available.  \citeone{GPG83} argue that almost all PN
are ionization bounded and have $\Mion$ continuously increasing with radius,
which would invalidate the Shklovsky method and its variations.
Alternatively, \citeone{Zij90} argues that, even when fully ionized, the
total mass of PN varies by a factor of ten, which would have the same
result.  This view is disputed by \citeone{Pot92}.

Recently, discussion on the validity of the Shklovsky method has
concentrated on PN in the galactic bulge.  Comparison of their Shklovsky
distances to the actual distance of the bulge should indicate the
reliability of the method and reveal any systematic effects.  STAS91
conclude that the agreement is acceptable, implying that the Shklovsky
method can be used to derive distances.  Based on almost the same sample,
PZ92 arrive at the opposite conclusion.  The difference between the two is
that the first paper uses optical diameters, which show a lack of compact
objects, while the second paper prefers radio diameters.  From the above
equations it can be seen that the diameter has the largest effect on the
calculated distance, compared to the other parameters.  For faint objects
there is also a discrepancy between the radio flux density and the H$\beta$
flux of up to a factor of two \cite{TAS92}, and the Shklovsky distances
depend on which flux is used.  However, this is a minor effect compared to
the diameter.

\subsection{Shklovsky distances to bulge planetary nebulae}

\Tabref{tab.Shklovsky} lists the calculated Shklovsky distances for our
sample of bulge PN\@.  We have assumed $\Mion=0.20$\,\msol\ and
$\varepsilon=0.5$, which are the same parameters as used by STAS91 and
PZ92.  Distances are shown for the two extreme cases, a sphere and a shell.
Almost all our distances are far beyond that of the galactic bulge
(\about7.7~kpc, \citebare{Rei89}), showing that the Shklovsky method does
not yield good distances for these objects.  We cannot do a statistical
analysis because our objects were chosen on the basis of a small expected
angular size.  However, since precisely these objects are the origin of the
discrepancy between the conclusions of STAS91 and PZ92, our results are a
strong vindication of the latter paper.

\begin{table}
\caption[]{Shklovsky distances and ionized masses\label{tab.Shklovsky}}
\begin{flushleft}
\begin{tabular}{
l                               c         c          c         c    c         c
       }
    \hline\noalign{\smallskip}
Object                  & \SGHz        & \mc{2}{c}{$d$ (kpc)} &
&\mc{2}{c}{\Mion\ ($10^{-3}$\msol)}\\
                        & (mJy)        & sphere     & shell   & & sphere  &
shell    \\
   \noalign{\smallskip}
    \hline\noalign{\smallskip}
321.0$+$03.9            & --           & \mc{5}{l}{(not a bulge PN)}
     \\ 
350.9$+$04.4            & 61           &   \z9.6    &    10.7   & &   117   &
\z88    \\ 
351.1$+$04.8            & 26           &    10.3    &    11.8   & &  \z97   &
\z68    \\ 
353.3$+$06.3            & 17           &    13.8    &    15.4   & &  \z46   &
\z35    \\ 
355.4$-$02.4            & 30           &            &   \z7.8\ra& &         &
197\ra \\ 
355.7$-$03.5            & 72           &    10.6    &    11.9   & &  \z90   &
\z67    \\ 
355.9$+$03.6            & 25           &    20.4    &    23.0   & &  \z17   &
\z13    \\ 
356.2$-$04.4            & 49           &    10.7    &    11.8   & &  \z89   &
\z69    \\ 
356.9$+$04.5            & 22           &    15.1    &    16.7   & &  \z37   &
\z29    \\ 
357.2$+$07.4            & 28           &    14.4    &    15.9   & &  \z42   &
\z32    \\ 
357.4$-$03.2            & 25           &    11.2    &    12.5   & &  \z79   &
\z60    \\ 
358.2$+$04.2            & 18           &    10.2    &    11.7   & &  \z99   &
\z70    \\ 
358.8$+$04.0            &\z3           &            &    10.1\ra& &         &
103\ra \\ 
359.8$+$03.7            & 18           &    13.7    &    15.2   & &  \z48   &
\z36    \\ 
000.1$+$17.2            & 19           & \mc{5}{l}{(probably not a bulge PN)}
     \\ 
000.2$-$01.9            & 14           &            &   \z6.1\ra& &         &
360\ra \\ 
003.1$+$03.4            & 10           &            &    11.3\ra& &         &
\z77\ra \\ 
004.6$+$06.0            & 15           &            &   \z9.0\ra& &         &
137\ra \\ 
005.0$+$04.4            & 18           &    19.7    &    21.8   & &  \z19   &
\z15    \\ 
006.4$+$02.0            & 59           &    10.5    &    11.7   & &  \z92   &
\z71    \\ 
008.2$+$06.8            & 13           &    21.1    &    23.3   & &  \z16   &
\z13    \\ 
\noalign{\smallskip}
\hline
\end{tabular}\\
$^a$calculated from the 10\%-contour diameter \\
\end{flushleft}
\end{table}

\subsection{Mass determinations}

The failure of the Shklovsky method to reproduce the correct distances for
these PN indicates that the assumption of a constant mass of 0.2\,\msol\ is
false.  Since we know the distance to the bulge, we can instead use
\eqref{eq.Shklovsky} to derive nebular masses.  The last two columns in
\tabref{tab.Shklovsky} show the results for the sphere and the shell models,
using $d=7.7$\,kpc.

\begin{figure}
\centerline{
\psfig{bbllx=10mm,bblly=42mm,bburx=195mm,bbury=156mm,%
figure=dummy,width=\the\hsize,scale=95}
}
\caption[]{
Mass-radius relationship for the galactic-bulge PN\@.
\label{fig.masses}}
\end{figure}

The masses we derive range from 0.013 to 0.4\,\msol.  They are plotted as
function of radius in \figref{fig.masses}, which can be compared to similar
figures in \citeone{P+A89} and \citeone{DHT90}.  Our objects fill in the
gap in the lower corner of the mass-radius diagram.  The most compact of
our objects has a radius of 0.017\,pc and, at a nominal expansion velocity
of 15\,km/s, would have an `expansion age' of about $10^3\,$yr.  Using the
diameters of the most compact bulge PN, the minimum time for the transition
between AGB and PN could be derived.  The present sample is too small for
this purpose, but the derived age of $10^3\,$yr can be taken as an upper
limit for the fastest-evolving objects.

\section{Conclusions}

We have measured H$\alpha$ diameters for 21 PN\@.  The nebulae are found to
be very compact, with angular diameters of 1--2$^{\prime\prime}$.  We show
how the derived diameters depend on the source model used for deconvolving
the seeing disk, which introduces an uncertainty of 10--15\% for the most
compact objects.

The final diameters are mostly in good agreement with radio measurements,
but are somewhat smaller than values found in earlier optical surveys.
Those optical diameters probably suffered from observational or
instrumental problems (guiding, non-linearity of photographic plates) which
prevented diameters close to the seeing limit from being determined.  Also,
some imaging may have been affected by inclusion of the [NII] lines, which
often come from outlying areas of the nebula.

On the basis of the new diameters, we have reinvestigated the controversy
regarding the Shklovsky method for distance determination.  For our
objects, we show that the Shklovsky method greatly overestimates the
distance and is not useful.  This agrees with the conclusions of PZ92 but
not with STAS91\@.  The latter paper used the older optical diameters; the
newly determined diameters do not support their conclusions.

\begin{acknowledgements}
The observations were made during a test program of the NTT remote-control
system and we are grateful to ESO for making this telescope time available.
We thank Gianni Marconi for assisting with the observations and Hans
Kjeldsen for valuable advice on acquiring and calibrating the CCD images.
Romano Corradi kindly provided CCD frames from the survey by
\citeone{SCM92}.  We also thank Agn\`es Acker for useful discussions and
the referee, Michael Dopita, for suggesting several improvements to the
paper.
\end{acknowledgements}

\bibliography{dummy}

\end{document}